\RequirePackage{fix-cm}
\documentclass[twocolumn]{svjour3}          
\smartqed  
\usepackage{graphicx}
\usepackage{epstopdf}
\usepackage{epsfig}
\usepackage{natbib}
\usepackage{color}
\usepackage{float}

\begin{document}

\title{Application of background-oriented schlieren (BOS) technique to a laser-induced underwater shock wave 
}
\subtitle{}


\author{Shota Yamamoto         \and
        Yoshiyuki Tagawa       \and		Masaharu Kameda
}


\institute{Y. Tagawa \at
              Department of Mechanical Systems Engineering, Tokyo University of Agriculture and Technology, Japan,  \\
              Tel.: +81-42-3887407\\
              Fax: +81-42-3887407\\
              \email{tagawayo@cc.tuat.ac.jp}           
}

\date{Received: date / Accepted: date}

\maketitle

\begin{abstract}
We build an ultra-high-speed imaging system based on the background-oriented schlieren (BOS) technique in order to capture a laser-induced underwater shock wave. 
This BOS technique is able to provide two-dimensional density-gradient field of fluid and requires a simple setup. 
The imaging system consists of an ultra-high speed video camera, a laser stroboscope, and a patterned background. 
This system takes images every 0.2 $\mu$s. Furthermore, since the density change of water disturbed by the shock is exceedingly small, the system has high spatial resolution $\sim$ 10 $\mu$m/pixel. 
Using this BOS system, we examine temporal position of a shock wave. 
The position agrees well with that measured by conventional shadowgraph, which indicates that the high-speed imaging system can successfully capture the instantaneous position of the underwater shock wave that propagates with the speed of about 1500 m/s. 
The local density gradient can be determined up to $O$(10$^3$ kg/m$^4$), which is confirmed by the gradient estimated from the pressure time history measured by a hydrophone.  


\end{abstract}
\section{Introduction}
\label{intro}
Laser-induced shock waves in water have been extensively investigated over several decades (e.g. \citet{bell1967laser,vogel1996shock,noack1998single,sankin2008focusing,lauterborn2013shock}).
The shock waves were applied in various technological fields such as laser medicine (\citet{hirano2002novel, tominaga2006application, sankin2005shock,sankin2008focusing}). 
\citet{tagawa2012highly} utilized a laser-induced shock wave to generate a highly-focused supersonic microjet, which can be applicable for novel needle-free injection devices (\citet{tagawa2013needle}). 
\citet{Turangan2013highly} confirmed numerically that the laser-induced shock wave plays a key role for generating a high-speed microjet. 

The quantification of the shock in experiments is, however, challenging due to its ultra-high velocity (about 1500 m/s) and small change of liquid density. 
Thus a desired measurement system has to have high spatial and temporal resolutions. 
 
In recent years, the background-oriented schlieren (BOS) technique has drawn attention as a novel measurement technique (\citet{dalziel2000whole,richard2001principle,meier2002computerized}). 
This technique provides two-dimensional information of material density with quite a simple setup.
The technique also has the great potential for measuring a wide range of flows.
The technique was applied to a steady supersonic air flow over a cone-cylinder model (\citet{venkatakrishnan2004density}), a large-scale imaging outside a laboratory with natural backgrounds (\citet{hargather2010natural,hargather2013background}), and the time evolution of the blast wave generated through a micro-explosion in the air (\citet{suriyanarayanan2012density,venkatakrishnan2013density}).
The three-dimensional density field of supersonic air flow around an asymmetric body was quantified using BOS technique with algebraic reconstruction technique (ART) (\citet{ota2011three}). 

To the best of the authors' knowledge, an application of BOS technique to an underwater shock wave has not been conducted.
Here we for the first time utilize BOS technique to measure the underwater shock wave.
Reminded that the shock has ultra-high velocity and is accompanied with little density change in water.
Thus we construct an ultra-high-speed BOS system with an ultra-high-speed video camera, a laser stroboscope, and a background. 
\citet{murphy2011piv} reported that particle image velocimetry (PIV) is applicable to visualize shock waves propagating in transparent polydimethylsiloxane, which implies that an underwater shock wave could be measured by BOS technique.

The structure of this paper is as follows. We describe BOS technique in section \ref{sec:BOS}, followed by measurement results in section \ref{sec:Reslts_Discussion}. 
Section \ref{Concl} concludes our work.
%
%
\section{Measurement methods}
\label{sec:BOS}
\subsection{Principle of BOS technique}
\label{sec:Principle_BOS}
\begin{figure}[h!]
\begin{center}
\includegraphics[width=0.45\textwidth]{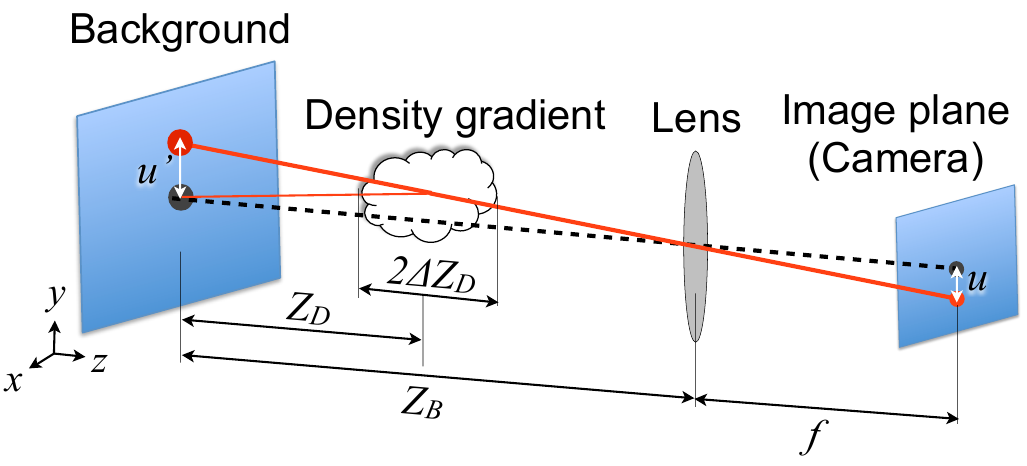}
\end{center}
\caption{Schematic of BOS technique.
A black dash line shows a ray from a background image without density gradients between the background and the image plane. 
A red solid line shows a ray with density gradients}
\label{fig:Schematic_BOS}       
\end{figure}

Figure \ref{fig:Schematic_BOS} shows a schematic of BOS technique. 
The principle of the technique is similar to conventional schlieren technique which exploits variation of refractive index of fluid due to its density gradients (the first-order differential value of density; $\nabla$$\rho$).
The variation of refractive index is obtained from two background images with random dots. 
One is a background image without density gradients between the background and the image plane (black dash line in Fig. \ref{fig:Schematic_BOS}). 
The other is a background image with density gradients (red solid line in Fig. \ref{fig:Schematic_BOS}). 
By using a PIV type cross-correlation algorithm, we can obtain background-element displacements.

\begin{figure}[h!]
\begin{center}
\includegraphics[width=0.45\textwidth]{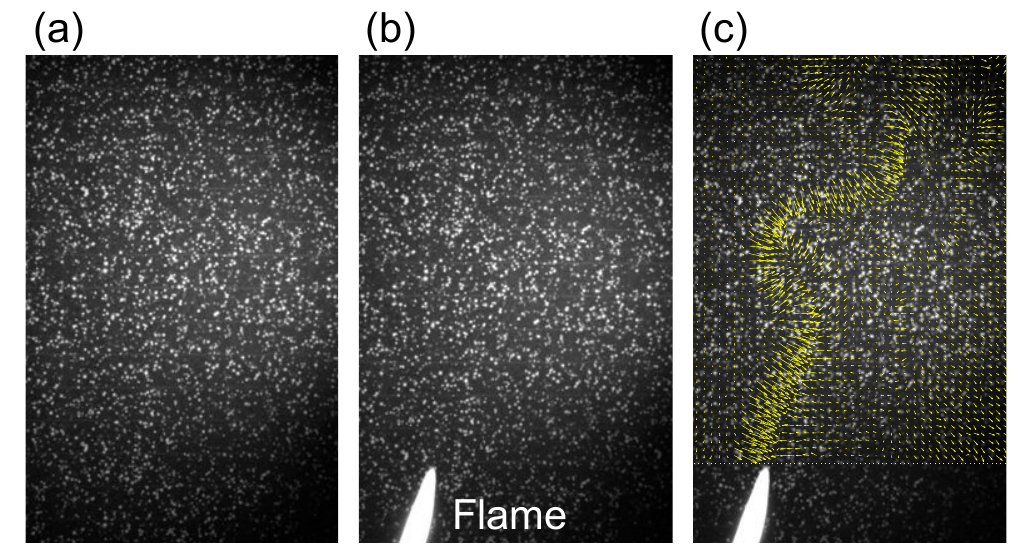}
\end{center}
\caption{Background images (a) without density gradients and (b) with density gradients due to a flame. (c) The image shows the difference between (a) and (b) using a PIV-type cross-correlation algorithm. Yellow vectors show the local displacements of the background}
\label{fig:Example_BOS}       
\end{figure}

Figure \ref{fig:Example_BOS} shows examples of the obtained images, in which thermal plume were captured by a still camera.
The undisturbed background image and the disturbed image are shown in Fig. \ref{fig:Example_BOS}(a) and (b), respectively. 
Figure \ref{fig:Example_BOS}(c) shows the field of background-element displacements shown as yellow vectors. 

These background-element displacements are related to density gradients as (\citet{venkatakrishnan2004density}):
\begin{equation}
\label{equation:disp_index}
v = \frac{Z_Df}{Z_B}\frac{1}{n_0}\int_{Z_D-\Delta Z_D}^{Z_D+\Delta Z_D}\frac{\partial n}{\partial y} dz
\end{equation}
\begin{equation}
\label{equation:Gladstone-dale}
n=K\rho+1,
\end{equation}
where $v$ is the background-element displacement in the vertical axis ($y$ direction, see Fig. \ref{fig:Schematic_BOS}), $Z_D$ is the distance from the background to the schlieren object, $f$ is the focal length, $Z_B$ is the distance from the background to the lens, $\Delta Z_D$ is one half of the thickness for schlieren object, $n$ is the refractive index, $z$ is the line-of-sight direction, $\rho$ is the density, and $K$ is the Gladstone-Dale constant.
Note that the background-element displacements obtained using BOS technique (Eq. \ref{equation:disp_index}) are related to the integrated value of the density gradients. 

%
\subsection{Experimental setup and conditions}
\label{sec:Setup}
\begin{figure}[h]
\begin{center}
\includegraphics[width=0.5\textwidth]{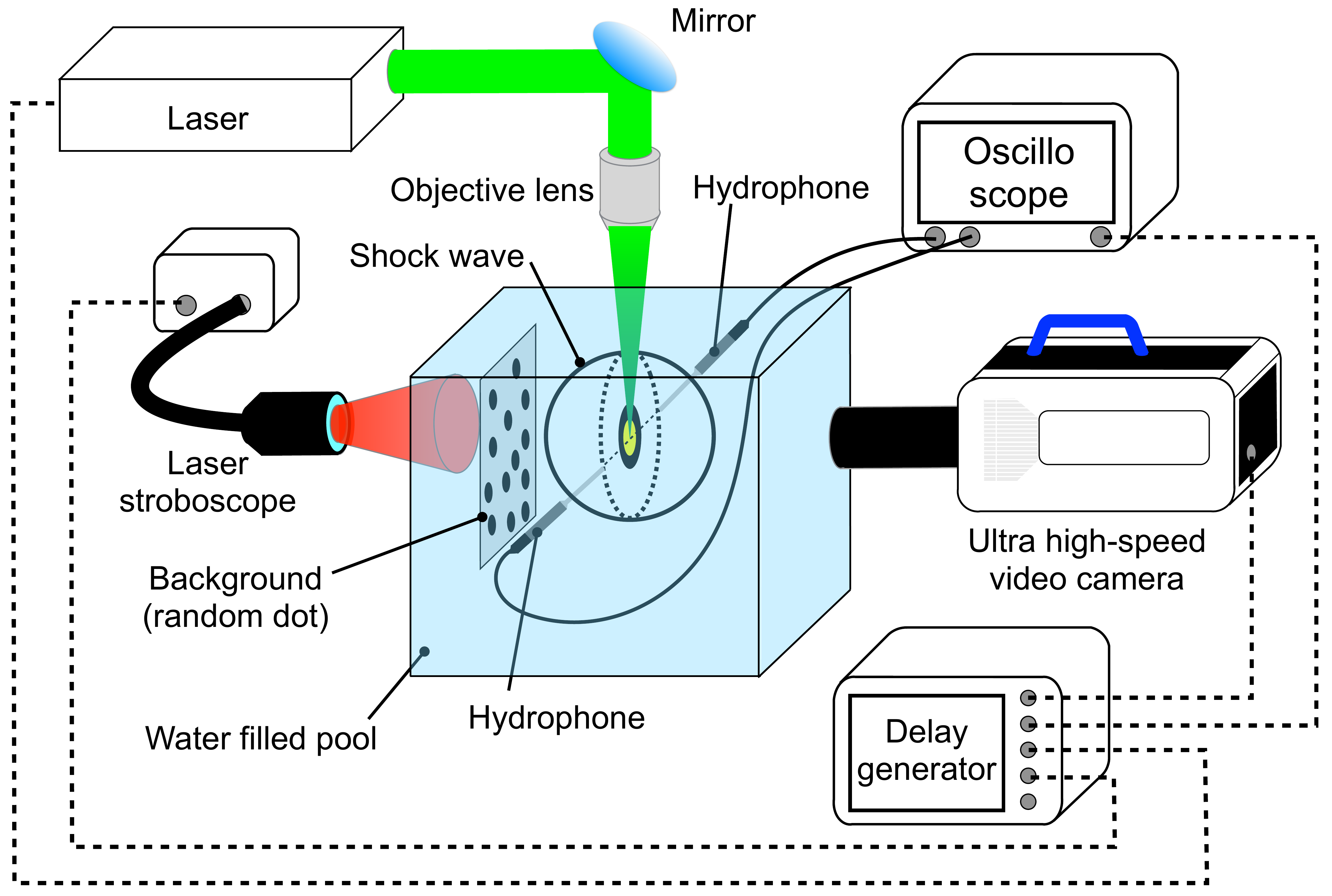}
\end{center}
\caption{Schematic of experimental setup}
\label{fig:Setup}       
\end{figure}

Figure \ref{fig:Setup} shows our experimental setup. 
Our BOS setup consists of a background with random dot pattern, a schlieren object (a density gradient), an ultra-high-speed camera, and a back light.
A 532 nm, 6 ns, laser pulse (Nd: YAGlaser Nano S PIV, Litron Lasers ltd., UK) is focused through a microscope objective lens (SLMPLN20x, magnification: 20$\times$, N.A.: 0.25, Olympus co., Japan) to a point inside a water-filled pool ($450\times300\times300$ mm), inside which a shock wave emerges. 
We record images using an ultra-high-speed video camera (Kirana, Specialized  Imaging co., UK) with up to $5\times10^6$ frames per second and 924$\times$768 pixel array. 
The experiment requires exceedingly short exposure time since the flow is non-stationary. 
Hence we utilize a laser stroboscope (CAVILUX Smart, CAVITAR co., Finland) with a pulse width of 20 ns as an illumination source.
Its repetition rate is also up to 5 MHz. 
A digital delay generator (Model 575, BNC co., USA) synchronizes the camera, the laser, and the stroboscope.

We obtain background-element displacements using a PIV-type cross-correlation algorithm. 
We use 50\% overlap with the Fast Fourier Transform (FFT) multi pass interrogation (from 32 pixel to 4 pixel).

For validating the high-speed BOS system, we  conduct two kinds of experiments. 

First, we compare temporal positions of a shock wave measured by BOS technique with that by shadowgraph, a common conventional method to visualize an underwater shock wave (\citet{bell1967laser,sankin2008focusing,brujan2008}).
Shadowgraph technique consists of a simple optical setup; a schlieren object (a density gradient), a camera, and a back light.
When we take shadowgraph images with our setup, we just remove the background from the high-speed BOS system. 
\begin{figure}[h]
\begin{center}
\includegraphics[width=0.45\textwidth]{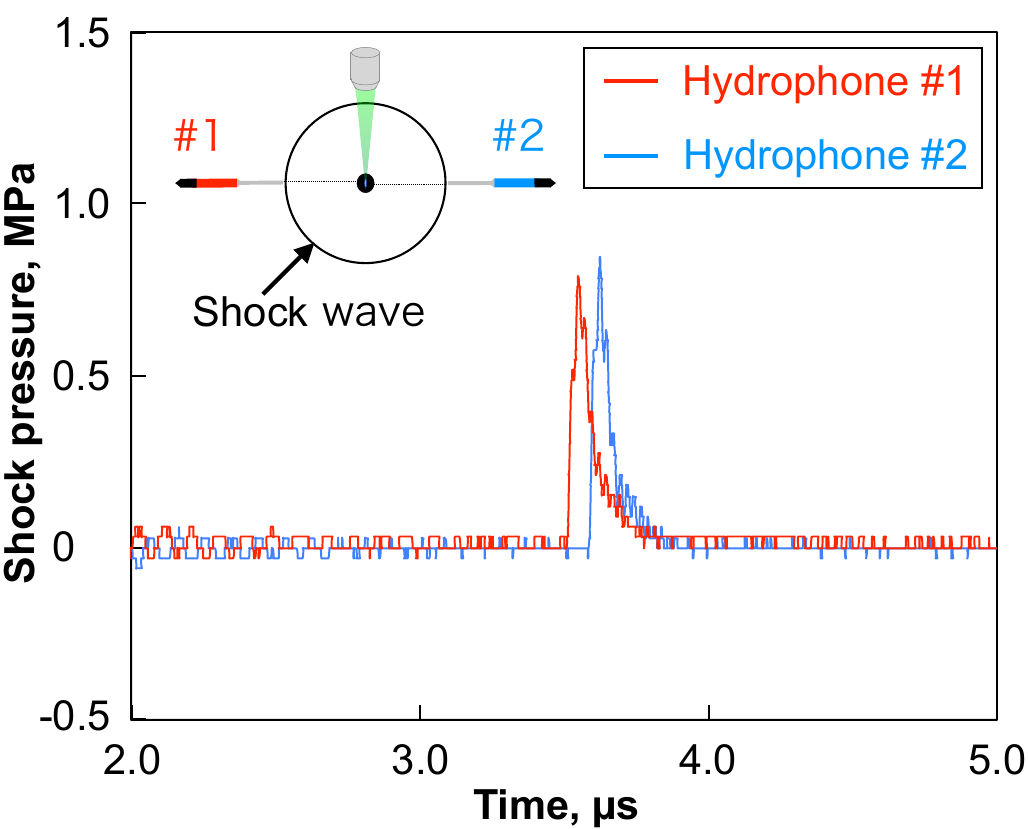}
\end{center}
\caption{Time history of shock pressures measured by two hydrophones arranged at right angles to the direction of a laser beam at a stand-off distance of 5 mm. Input laser energy is 1.9 mJ}
\label{fig:hyd_shock}       
\end{figure}
\begin{figure}[h]
\begin{center}
\includegraphics[width=0.45\textwidth]{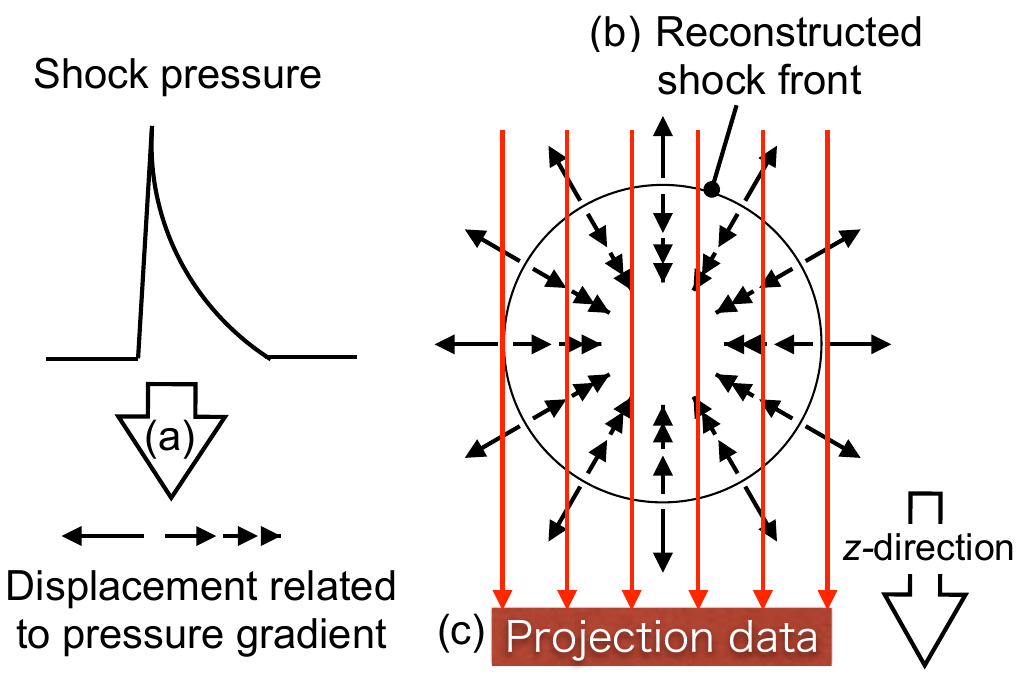}
\end{center}
\caption{Procedure of converting shock pressure to displacements. 
(a) Conversion of shock pressure measured by a hydrophone to displacements related to pressure (density) gradients. 
(b) Reconstructed displacement field with assuming an axisymmetric shock structure. 
(c) Projection of the reconstructed displacement field toward $z$-direction}
\label{fig:procedure}       
\end{figure}
Note that shadowgraph records local brightness that corresponds to second derivative of the fluid density $\nabla^2$$\rho$ (\citet{gs2001schlieren}).
Since the displacement obtained from BOS technique is related to the density gradient $\nabla$$\rho$, we differentiate the density gradient for obtaining second derivative of the fluid density $\nabla^2$$\rho$.
In order to extract the mean shock position we approximate the shape of the shock as a sphere (a circle in two-dimensional projection). 

Second, we verify the displacements obtained from our BOS technique by comparing them with those estimated from hydrophones. 
We use two hydrophones (Muller-Platte Needle Probe, Muller ins., Germany) and the high-speed BOS system simultaneously.
Each hydrophone is arranged at right angles to the direction of a laser beam at a stand-off distance of about 5 mm and is connected to an oscilloscope (ViewGo II DS-5554A, Iwatsu co., Japan).
Figure \ref{fig:hyd_shock} shows typical hydrophone measurements.
Both shock pressures are almost the same, which indicates that the shock wave has an axisymmetric structure.
The small difference is due to different positions of hydrophones; a stand-off distance of the hydrophone $\#1$ is 5.14 mm and the hydrophone $\#2$ is 5.27 mm.

A procedure for estimating displacements from hydrophone measurements is as follows:
We first calculate pressure gradient from pressure time history (Fig. \ref{fig:procedure}(a)).
We obtain the refractive index using Gladstone-Dale equation (Eq. \ref{equation:Gladstone-dale}) and Tait equation;
\begin{equation}
\label{equation:Tait}
\frac{p+B}{p_0+B}=\left(\frac{\rho}{\rho_0}\right)^m,
\end{equation}
where $p_0$ is the hydrostatic pressure, $B$ and $m$ are constants based on physical properties.
In the case of water, $B$ is 314 MPa and $m$ is 7.
The calculated variation of refractive index is then related to displacements via Eq. \ref{equation:disp_index}.
We then reconstruct the two-dimensional displacement field on $x$--$z$ plane from hydrophone results as shown in Fig. \ref{fig:procedure}(b) since a hydrophone provides point-wise pressure information while the displacement obtained from BOS technique is a value integrated along  $z$-axis as shown in Fig. \ref{fig:procedure}(c) (cf. Fig. \ref{fig:Schematic_BOS}).
Reminded that we here assume an axisymmetric structure of a laser-induced shock wave, which is verified from hydrophone measurements in Fig. \ref{fig:hyd_shock}.
Finally,  we obtain the estimated displacements by integrating the reconstructed displacement fields along $z$-axis.
%

We set two levels of laser energy: 1.9 mJ and 3.0 mJ.
In addition, we change spatial resolutions of images, which depend on the focal length $f$ and the distance from the background to the lens $Z_B$.
With higher spatial resolution of 8.4 $\mu$m/pixel, $f$ and $Z_B$ are respectively 360 mm and 120 mm, while with lower spatial resolution of 14.5 $\mu$m/pixel, $f$ and $Z_B$ are respectively 250 mm and 120 mm. 
%

%
%

\begin{figure*}
\begin{center}
\includegraphics[width=1\textwidth]{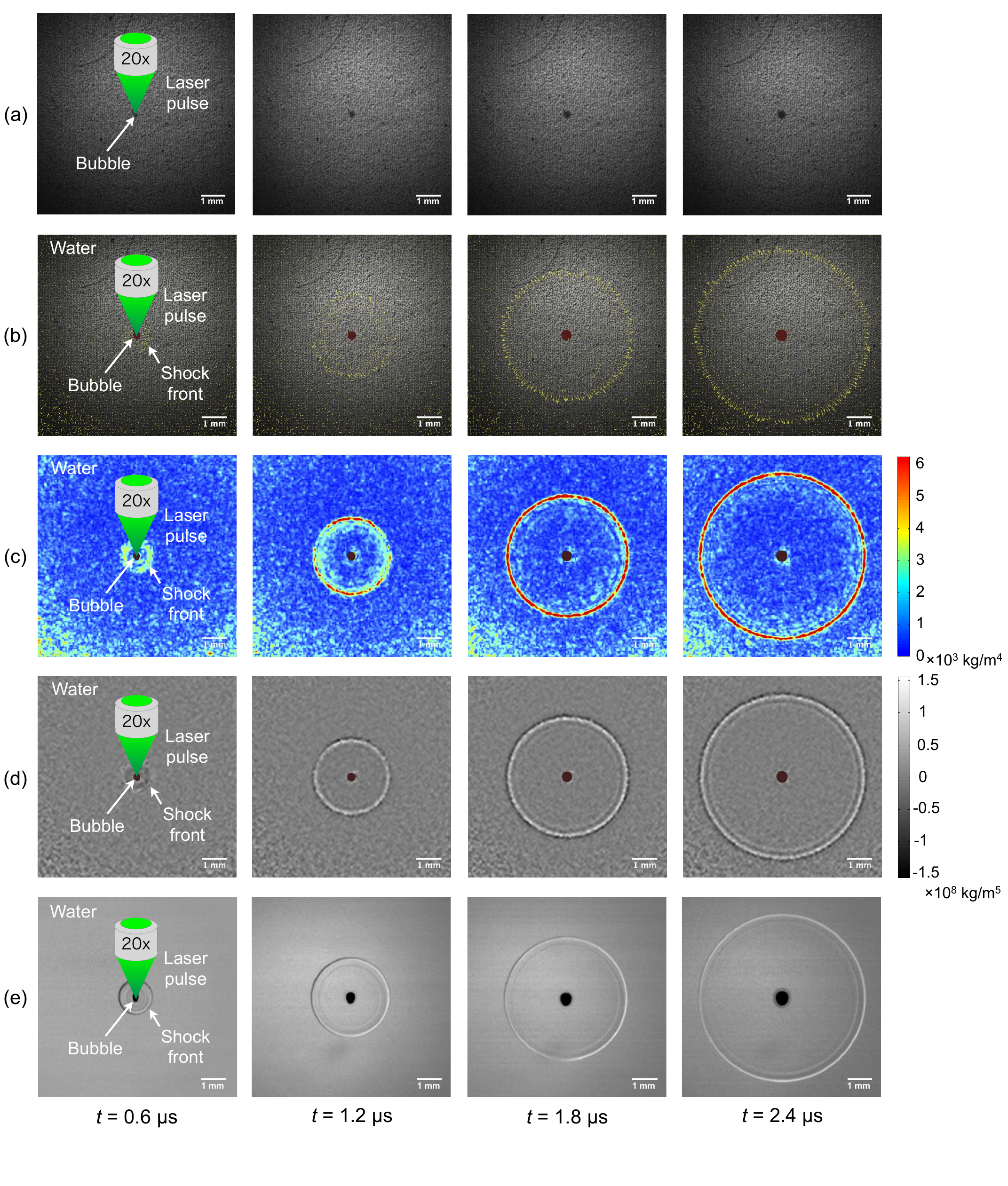}
\end{center}
\caption{The spatiotemporal evolution of the laser-induced shock wave after the laser pulse beam being fired. 
The laser pulse beam is illuminated from the top of these images.
(a)--(e) are shown at the same timing.
We mask the region of bubble with brown elliptic area in (b)--(d).  
(a) Original images obtained from the ultra-high-speed recording system.
(b) Displacement field obtained by BOS technique.
The displacements are shown as yellow vectors.
The displacement corresponds to the density gradient ($\nabla$$\rho$).
(c) Magnitude of the density gradient field ($|\nabla$$\rho|$).
(d) The second-order differential value of density ($\nabla^2$$\rho$) calculated from the displacement values (b).
(e) The shock wave visualized by shadowgraph.
A dark streak pattern due to shock wave shows the second-order differential value of density ($\nabla^2$$\rho$)
}
\label{fig:scries}       
\end{figure*}

\section{Results and Discussion}
\label{sec:Reslts_Discussion}
%
\subsection{Position of an underwater shock wave visualized by BOS technique and shadowgraph}
\label{sec:Visualization} 

We show spatiotemporal evolution of a shock wave measured by BOS technique and shadowgraph in Fig. \ref{fig:scries}.
Original images obtained from the high-speed BOS system are displayed in Figure \ref{fig:scries}(a).
Figure \ref{fig:scries}(b) shows the vector field of displacements of background-elements, which correspond to the density gradients.
The regions of non-zero density gradients (non-zero vectors) radially propagates from a laser-induced bubble.
For clear presentation, the magnitudes of the displacements are shown in Fig. \ref{fig:scries}(c).
Red color shows the largest displacement.
We observe spherical-symmetric distribution of the shock as reported in many previous researches (e.g. \citet{vogel1996shock,petkovvsek2005optodynamic,brujan2008}).  
Figure \ref{fig:scries}(d) shows the second-order differential value of density ($\nabla^2$$\rho$) obtained by differentiating the density gradient filed ($\nabla$$\rho$).  
For comparison, a shock wave in the same condition visualized  by shadowgraph is shown in Fig. \ref{fig:scries}(e).
A dark streak pattern due to a shock wave shows the second-order differential value of density ($\nabla^2$$\rho$).
Both Fig. \ref{fig:scries}(d) and (e) look almost the same, showing sphere-like shapes of the shocks.

\begin{figure}[h!]
\begin{center}
\includegraphics[width=0.45\textwidth]{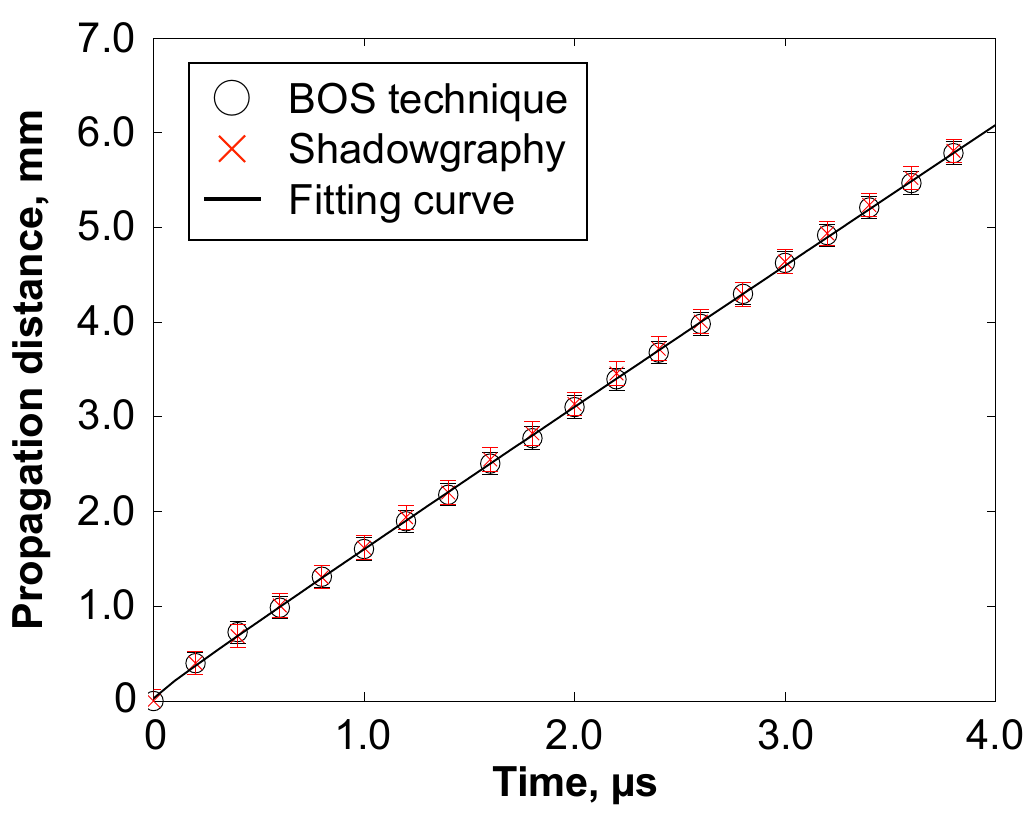}
\end{center}
\caption{Propagation of the shock front plotted as a function of  time. The solid line presents the fitting curve proposed by \citet{Dewey2001}}
\label{fig:Position_Velocity}       
\end{figure}

We here quantitatively compare shock positions by both methods.
We plot both shock radii, namely mean shock positions, as a function of time in Fig. \ref{fig:Position_Velocity}.  
The temporal evolution of the shock position from the BOS system agrees with that by shadowgraph, within measurement error $\pm$120 $\mu$m.

For a spherical shock wave, the temporal evolution can be described as (\citet{Dewey2001}),
\begin{equation}
\label{equation:Shock_radius}
R_s=C_1+C_2a_0t+C_3\ln(1+a_0t)+C_4\sqrt{\ln(1+a_0t)},
\end{equation}
where $R_s$ is the fitted shock position, $C_1$, $C_3$, and $C_4$ are the fitting constants, $C_2$ is set to 1, $a_0$ is the speed of sound in water ($a_0$=1483 m/s), and $t$ is an elapsed time after the laser being fired.
It nicely fits the data, indicating that the shock visualization is valid.
Thus the high-speed BOS system is able to detect the instantaneous position of the non-zero density gradient, i.e. an underwater shock wave.

%

%

\subsection{Comparison of BOS technique with hydrophone measurement}
\label{sec:Quantification}
\begin{figure}[h!]
\begin{center}
\includegraphics[width=0.45\textwidth]{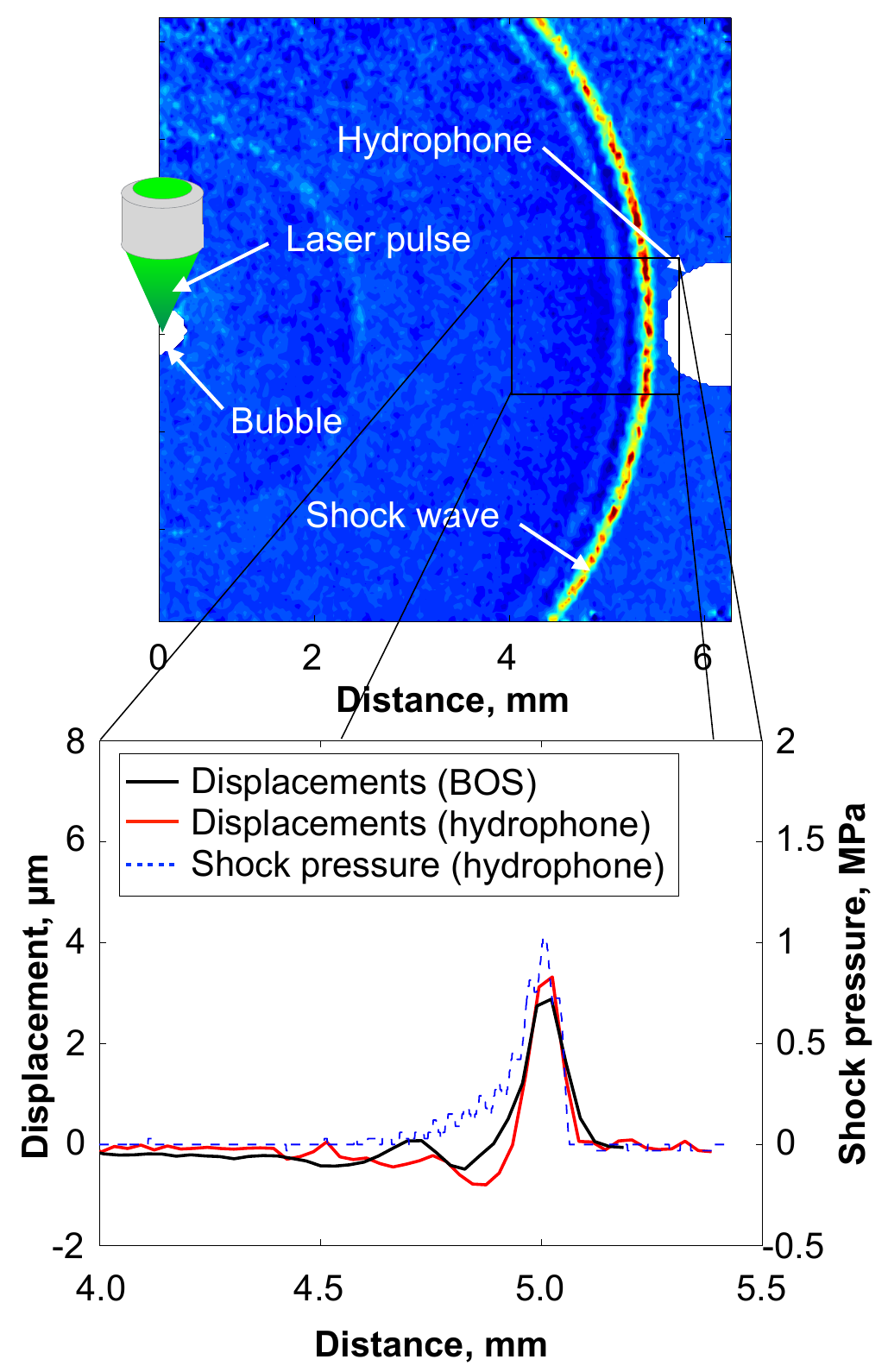}
\end{center}
\caption{Displacements obtained from BOS technique and those calculated from hydrophone measurements with 1.9 mJ of laser energy.
Upper figure presents displacement fields of the $x$-direction ($u$) obtained from BOS technique with spatial resolution of 8.4 $\mu$m/pixel.
Lower figure compares displacements.
The black line represents displacements obtained from BOS technique; the red line shows displacements calculated from hydrophone measurements; the blue dashed line is shock pressure measured by a hydrophone}
\label{fig:comp_peakp_low}
\end{figure}

\begin{figure}[h!]
\begin{center}
\includegraphics[width=0.45\textwidth]{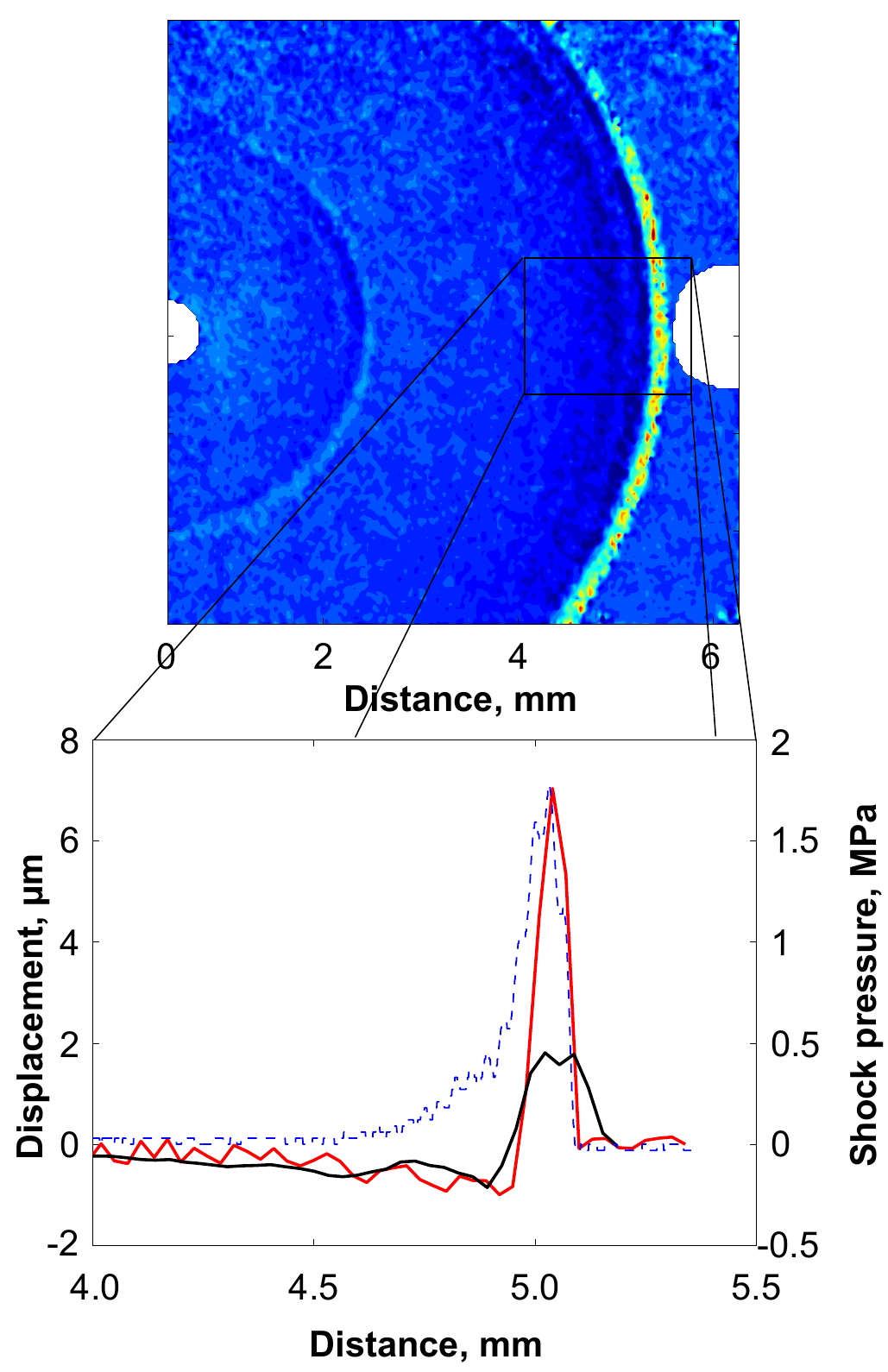}
\end{center}
\caption{Upper panel: Magnitude of the density gradient field presented in the same manner for Fig. \ref{fig:scries}(c). Lower panel: Displacements obtained from BOS technique and those calculated from hydrophone measurements. The laser energy is 3.0 mJ.
Lines in this figure represent the same quantities as those in Fig. \ref{fig:comp_peakp_low}
}
\label{fig:comp_peakp_high}
\end{figure}

In this section we compare the displacements obtained from BOS technique with those calculated from hydrophone measurements.
The displacement fileds of $x$-components ($u$) for two different laser energies are shown in the upper side of Figs. \ref{fig:comp_peakp_low} and \ref{fig:comp_peakp_high}.
The spatial resolution is 8.4 $\mu$m/pixel.
The shape of the shocks are almost spherical for all the laser energies.
For quantitative discussion, the displacements $u$ are shown in the lower side of Figs. \ref{fig:comp_peakp_low} and \ref{fig:comp_peakp_high}, where the displacements obtained from BOS technique and a hydrophone are respectively presented as a black line and a red line.
Shock pressure measured by a hydrophone is also displayed as a blue-chain line for reference.


For lower laser energy (1.9 mJ), the maximum displacement obtained from BOS technique fairly agree with that from hydrophone measurements.
In this particular case the maximum density gradient is 6$\times$10$^3$ kg/m$^4$ based on peak pressure of 1 MPa and shock width of 60 $\mu$m.

As the laser energy increases, however, the maximum displacement obtained by BOS technique is considerably smaller than the displacement calculated from hydrophone measurements. 
The ratios of the maximum displacement from BOS technique to that from hydrophones are 87\% and 26\% for laser energy 1.9 mJ and 3.0 mJ, respectively.
It indicates that in the cases for higher laser energy, the largest density gradient exceeds a measurable limit of the high-speed BOS system.
For input laser energy 3.0 mJ, the maximum density gradients are 1.4$\times$10$^4$ kg/m$^4$.
Therefore current high-speed BOS system can be applicable to a flow field, density gradient of which is of the order of 10$^3$ kg/m$^4$. 

\begin{figure}[h!]
\begin{center}
\includegraphics[width=0.45\textwidth]{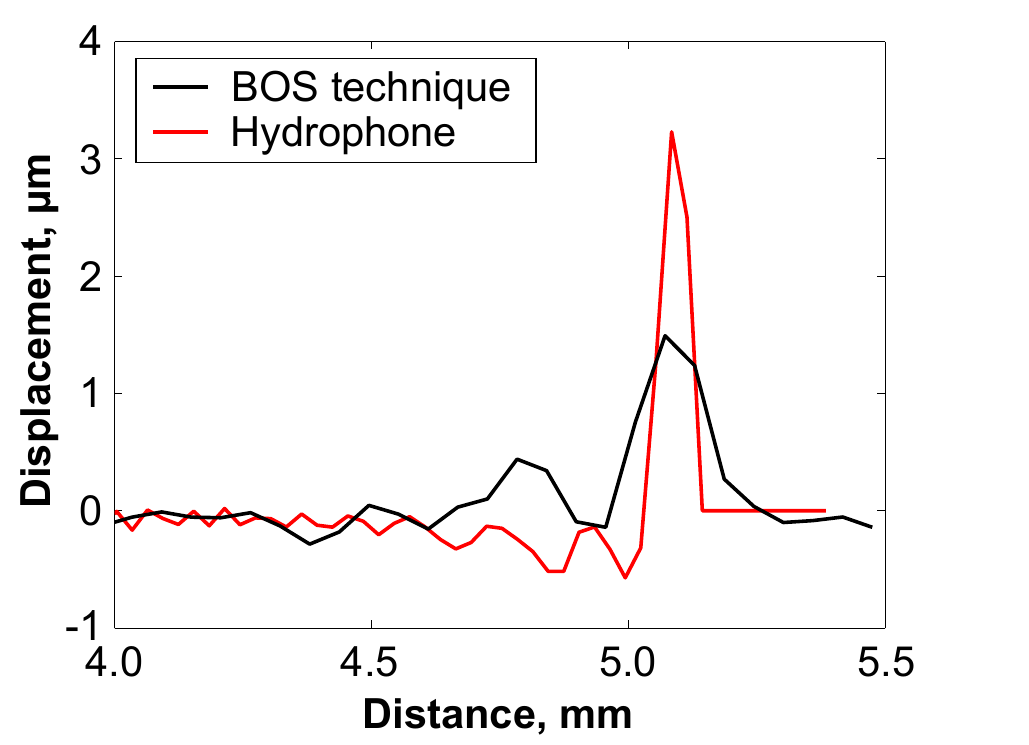}
\end{center}
\caption{Displacements obtained from BOS technique and those calculated from hydrophone measurements with 14.5 $\mu$m/pixel of spatial resolution. 
Input laser energy is 1.9 mJ.
The black line represents displacements obtained from BOS technique; the red line shows displacements calculated from hydrophone measurements}
\label{fig:comp_resol}
\end{figure}

In order to verify the effect of spatial resolution of our BOS system, we measure the underwater shock wave with lower resolution (14.5 $\mu$m/pixel)  in the condition of 1.9 mJ laser energy. 
Measured displacements are shown in Fig. \ref{fig:comp_resol}.
The maximum displacement for 14.5 $\mu$m/pixel is 46\% of that from a hydrophone, which is obviously less than the value obtained with higher spatial resolution (8.4 $\mu$m/pixel). 
We thus confirm that the applicable range of the BOS system depends on its spatial resolution as pointed out by \citet{goldhahn2007background}.

%
\section{Conclusion}
\label{Concl}
We have applied background-oriented schlieren (BOS) technique to an underwater shock wave.
Challenges in the measurements are ultra-high velocity of the shock wave in water and small density fluctuation of water.
We have constructed an ultra-high speed BOS system that is able to take images every 0.2 $\mu$s with high-spatial resolution $\sim$ 10 $\mu$m/pixel.
For its validation, we have measured the shock by shadowgraph or hydrophones simultaneously with the BOS system.

We have found that the spatiotemporal position of the shock detected by BOS technique has agreed with that by shadowgraph.
It indicates that the high-speed BOS system is able to detect instantaneous position of the shock wave that changes local density of water.
In addition, in order to estimate a measurable range for density gradient we have compared the displacements obtained from BOS technique with those from hydrophones.
We have varied input laser energy and spatial resolution of the BOS system. 
The measurable range increases with an increase of spatial resolution of the BOS system.
It has been found that with current high-speed BOS system the density gradient of the order of 10$^3$ kg/m$^4$ can be reasonably measured. 
Spatial resolution higher than 8.4 $\mu$m/pixel might enable us to measure a flow field with even higher density gradients. 


\begin{acknowledgements}
This work was supported by JSPS KAKENHI Grant Number 26709007.
We thank Keisuke Hayasaka for his assistance with the experiments.
\end{acknowledgements}

\bibliographystyle{spbasic}      
\bibliography{BOS}   

\end{document}